\documentclass[10 pt,final,journal,letterpaper,oneside,twocolumn]{IEEEtran}
\usepackage{multicol}   
\usepackage{multirow}
\usepackage[pdftex]{graphicx}
\usepackage[ruled]{algorithm2e}
\usepackage{algorithmic}
\usepackage[small]{caption}
\usepackage{float}
\usepackage{amssymb,amsmath}
\usepackage{graphicx}
\usepackage{subfigure}

\usepackage{xspace}
\usepackage{fancyhdr}
\usepackage{amssymb,amsmath}
\usepackage{textcomp} 
\usepackage{epstopdf}
\usepackage{bbding}
\usepackage{cite} 
\usepackage{color}
\usepackage{multicol}
\usepackage{multirow}
\usepackage{array}
\usepackage{pifont}

\begin{document}
%
\title{Integration of NOMA with Reflecting Intelligent Surfaces: A Multi-cell Optimization with SIC Decoding Errors}
\author{Wali Ullah Khan, \textit{Member, IEEE,} Eva Lagunas, \textit{Senior Member, IEEE,} Asad Mahmood, Zain Ali, Muhammad Asif, Symeon Chatzinotas, \textit{Fellow, IEEE,} and Bj\"orn Ottersten, \textit{Fellow, IEEE} \thanks{This work was supported by the Luxembourg National Research Fund (FNR) under the CORE project RISOTTI, grant C20/IS/14773976. \textit{(Corresponding author: Wali Ullah Khan)} 

Wali Ullah Khan, Eva Lagunas, Asad Mahmood, Symeon Chatzinotas, and Bj\"orn Ottersten are with SnT, University of Luxembourg (Emails: \{waliullah.khan, eva.lagunas, asad.mahmood, symeon.chatzinotas, bjorn.ottersten\}@uni.lu). 

Zain Ali is with Department of Electrical and Computer Engineering, University of California, Santa Cruz, USA (email: zainalihanan1@gmail.com).

Muhammad Asif is with Guangdong Key Laboratory of Intelligent Information Processing, College of Electronics and Information Engineering, Shenzhen Univerisyt, Shenzhen, Guangdong China (e-mail: masif@szu.edu.cn).

}}%

\markboth{IEEE Transaction on Green Communications and Networking, vol. x, no. y, ABC 2023}%
{Shell \MakeLowercase{\textit{et al.}}: Bare Demo of IEEEtran.cls for IEEE Journals} 

\maketitle

\begin{abstract}
Reflecting intelligent surfaces (RIS) has gained significant attention due to its high energy and spectral efficiency in next-generation wireless networks. By using low-cost passive reflecting elements, RIS can smartly reconfigure the signal propagation to extend the wireless communication coverage. On the other hand, non-orthogonal multiple access (NOMA) has been proven as a key air interface technique for supporting massive connections over limited resources. Utilizing the superposition coding and successive interference cancellation (SIC) techniques, NOMA can multiplex multiple users over the same spectrum and time resources by allocating different power levels. This paper proposes a new optimization scheme in a multi-cell RIS-NOMA network to enhance the spectral efficiency under SIC decoding errors. In particular, the power budget of the base station and the transmit power of NOMA users while the passive beamforming of RIS is simultaneously optimized in each cell. Due to objective function and quality of service constraints, the joint problem is formulated as non-convex, which is very complex and challenging to obtain the optimal global solution. To reduce the complexity and make the problem tractable, we first decouple the original problem into two sub-problems for power allocation and passive beamforming. Then, the efficient solution of each sub-problem is obtained in two-steps. In the first-step of For power allocation sub-problem, we transform it to a convex problem by the inner approximation method and then solve it through a standard convex optimization solver in the second-step. Accordingly, in the first-step of passive beamforming, it is transformed into a standard semi-definite programming problem by successive convex approximation and different of convex programming methods. Then, penalty based method is used to achieve a Rank-1 solution for passive beamforming in second-step. Numerical results demonstrate the benefits of the proposed optimization scheme in the multi-cell RIS-NOMA network.
\end{abstract}


\IEEEpeerreviewmaketitle

\section{Introduction}
\IEEEPARstart{R}{esearchers} throughout the world are looking towards beyond 5G technologies now that 5G wireless networks are becoming commercially available. As the need for wireless networks continues to rise and the limitations of 5G networks become more apparent, a technologically-driven paradigm shift towards beyond 5G networks is becoming increasingly important \cite{dang2020should}. Meanwhile, beyond 5G networks will significantly enhance and broaden 5G's application space. Beyond 5G networks have a peak throughput of up to 10 Tbit/s, allow for a connection density of 107 devices/km2, and are 100 times more energy efficient than 5G networks \cite{zhang20196g}. To provide adequate services for the forthcoming advanced information society, beyond 5G networks will need to significantly increase their scalability, adaptability, and efficiency by incorporating new technologies.

As the wireless industry progresses beyond 5G networks, many new technologies will be introduced, and these new requirements for future networks will have an impact on the key technology trends in its evolution process. In this regard, several technologies are expected to be the enablers for next-generation wireless networks. Among them, reflecting intelligent surfaces (RIS) \cite{9122596} and non-orthogonal multiple access (NOMA) \cite{9468352} are promising technologies. The RIS can smartly reconfigure the wireless signal between the transmitter and receiver using the passive elements. According to its working principle, RIS modifies the amplitude and phase of the reflective signal. By using this amazing feature, RIS can significantly enhance the performance of wireless communication in terms of coverage extension, user fairness, and channel secrecy \cite{opportunities}. On the other hand, NOMA has proved to be one of the key air interface techniques for high spectral efficiency and massive connectivity. This is because NOMA supports multiple users over the single spectrum resource simultaneously \cite{8357810}. This is possible due to superposition coding adopted by the transmitter and the receiver's successive interference cancellation (SIC) decoding process.

RIS is a new frontier of wireless communication technology that has gained significant attention in recent years \cite{badheka2023accurate}. The RIS system consists of many small, passive and inexpensive elements that can be used to reflect and manipulate electromagnetic waves in a controllable manner \cite{khan2022intelligent}. Moreover, IRS systems provide an innovative approach to improving the performance of wireless communication systems by enhancing signal strength, coverage, and security. These intelligent surfaces can be integrated into various communication systems such as unmanned aerial vehicles \cite{peng2023energy}, intelligent transportation systems \cite{khan2022opportunities}, satellite communications \cite{xu2021intelligent} and the Internet of Things \cite{zhu2021resource}. RIS technology has several advantages over traditional relay networks in several aspects. For example, it offers high spectral and energy efficiency, low complexity, and low power consumption \cite{bjornson2019intelligent}. Furthermore, it enables communication systems to operate in harsh and challenging environments, such as underground tunnels and buildings with thick walls. 

Various problems in RIS-NOMA systems using single-cell scenarios have been extensively studied in the literature. For example, Zuo {\em et al.} \cite{zuo2020resource} have investigated a joint resource optimization problem of the channel assignment, decoding order, power allocation and reflection matrix design to enhance the throughput of RIS-NOMA system. The study of \cite{mu2020exploiting} has jointly optimized the sum rate of the RIS-NOMA system through active and passive beamforming at BS and RIS, respectively. Basharat {\em et al.} \cite{basharat2022intelligent} have maximized the sum rate of the multi-carrier RIS-NOMA system by optimizing channel assignment, power allocation and RIS elements. The authors of \cite{li2021energy} have optimized the total power consumption of RIS-enabled mobile edge computing NOMA network by efficiently allocating the transmit power, passive reflection design, and decoding order. Guo {\em et al.} \cite{guo2020intelligent} have proposed two-phase shifter algorithms for the RIS-NOMA system to maximize the average sum rate subject to transmit power, minimum rate and reflection coefficient. Moreover, the research work in \cite{zhong2022mobile} has employed deep learning approach for resource optimization to improve the sum rate of the RIS-NOMA system. Of late, Ihsan {\em et al.} \cite{ihsan2022energy} have optimized passive and active beamforming along with efficient power allocation to maximize the energy efficiency of the RIS-NOMA system.

\subsection{Recent Advances in RIS-enabled multi-cell networks}
The performance of RIS in multi-cell wireless systems adopting orthogonal multiple access (OMA) has been extensively studied in the literature. For instance, in \cite{buzzi2021ris}, the authors examined a RIS-assisted multi-user wireless network and presented techniques for allocating resources in both single-cell and multi-cell deployments. Both scenarios were investigated, and the signal strength was enhanced using beamforming and phase shift design based on the estimated channel.
The authors of \cite{hashida2020intelligent} used RIS to extend wireless coverage in aerial networks. Their goal was to maximize network capacity while reducing inter-cell interference. When the best position to install RIS are selected, their results show that the proposed network with RIS outperforms the benchmark system without RIS. Moreover, the work in \cite{yu2022resource} presented an efficient resource allocation framework for reducing the total system load of RIS-enabled wireless networks. Their optimization problem was multi-variable and non-convex, resulting in high complexity. Using the majorization-minimization method, they were able to obtain a simple local solution that was also efficient. The work in \cite{cai2022irs} proposed a joint beamforming problem at the base station (BS) and RIS to reduce power consumption and enhance the sum capacity of the system. They adopted popular second-order cone programming, Riemannian manifold, minimization-majorization, weighted minimum mean square error, and block coordinate descent methods to solve the formulated problem. Furthermore, the authors of \cite{taghavi2022joint} maximized the sum capacity of RIS-enabled wireless networks through joint optimization of phase shift design and user association. They used alternating optimization first to handle the complicated con-convex problem and then exploited fractional programming and successive convex approximation methods to solve the sub-problems.

In addition, RIS has also been integrated into multi-antenna wireless networks. For example, the work of \cite{pan2020multicell} optimized the weighted sum rate of all users while satisfying the power constraint and unit modulus constraint imposed by each BS. Due to the non-convex optimization problem, they first transformed it into a convex one and then used a block coordinate descent approach to iteratively optimize the precoding matrices and phase shifts. Huang {\em et al.} \cite{huang2019reconfigurable} have proposed an energy-efficient optimization framework in multi-user RIS-enabled networks. In particular, they explored a non-convex transmit power and phase shift design problem through alternating optimization. Xu {\em et al.} \cite{xu2022deep} have provided deep reinforcement learning for joint beamforming design in RIS-enabled wireless networks. Qiu {\em et al.} \cite{qiu2022joint} have enhanced the sum capacity of multi-cell RIS network by joint beamforming at BSs and RIS. The researchers in \cite{xie2020max} allocated fair system resources to maximize the minimum weighted capacity of the RIS-enabled multi-antenna network. They solved the transmit beamforming and phase shift design problem using alternative optimization. The authors of \cite{luo2021reconfigurable} maximized the ergodic weighted sum capacity of RIS-enabled wireless networks under imperfect channel information. They employed fractional programming and complex circle manifold methods to design beamforming and phase shift matrices. Moreover, the work in \cite{rezaei2022energy} proposed an energy-efficient system design for RIS-enabled wireless networks. They adopted the Dinkelbach method for efficient antenna selection and beamforming to reduce overall energy consumption. The paper in \cite{al2021performance} maximized the system throughput of RIS-enabled multi-antenna networks. They proposed different combining and power-assigning techniques in all cells. Furthermore, the authors of \cite{cai2021intelligent} designed transmit beamforming and phase shift design to reduce the energy consumption of RIS-enabled wireless networks. Late, researchers in \cite{zhang2022meta} considered active RIS for indoor communication capable of signal transmission and reflection at both sides. They designed hybrid beamforming to enhance the sum capacity of the system.
\begin{table*}[tbp]
\centering
\caption{Comparison of the proposed work with existing RIS-enabled multi-cell works.}
\label{Rel_Works}
\scriptsize
\begin{tabular}{| m{2em} | m{1.2cm}| m{1.3cm} |  m{2.5cm} | m{1.5cm}| m{5cm} | m{2cm} |} 
  \hline
  	\hline
  \textbf{Ref.} & \textbf{Cell(s)} & \textbf{IRS(s)} & \textbf{Work Objective} &\textbf{OMA/NOMA} & \textbf{Proposed Solution} & \textbf{SIC Assumption}\\
  \hline \hline
  \cite{buzzi2021ris} & Two-cells & Single-RIS & Improving SNR/SINR & OMA & Optimizing BS transmit power and RIS configuration design & ------ \\
   \hline
 \cite{hashida2020intelligent} & Two-cells & Two-RIS & Improving SINR  & OMA & Optimizing RIS placement & ------ \\
    \hline
\cite{yu2022resource} & Multi-cell & Multi-RIS & Minimizing resource consumption & OMA & Optimizing reflection coefficient at RIS  & ------ \\
	\hline
\cite{cai2022irs}& Multi-cell & Single-RIS & Reducing total power consumption & OMA & Optimizing beamforming at BS and phase shist at RIS & ------ \\
   \hline
\cite{taghavi2022joint}& Multi-cell & Multi-RIS & Maximizing total spectral efficiency & OMA & Optimizing user association and phase shift at RIS & ------ \\
   \hline
\cite{pan2020multicell}& Two-cell & Single-RIS & Maximizing weighted sum rate & OMA & Optimizing precoding at BS and phase shift at RIS & ------ \\
   \hline
\cite{huang2019reconfigurable}& Single-cell & Single-RIS & Maximizing energy efficiency & OMA & Optimizing transmit power of BS and phase shift at RIS & ------ \\
     \hline
\cite{xu2022deep}& Single-cell & Single-RIS & Maximizing average achievable rate & OMA & Beamforming at BS and RIS & ------ \\
   \hline
\cite{qiu2022joint}& Multi-cell & Single-RIS & Maximizing sum rate & OMA & Optimizing beamforming at BS and phase shift at RIS & ------ \\
   \hline
\cite{xie2020max}& Multi-cell & Single-RIS & Max-min SINR & OMA & Optimizing transmit beamforming at BS and reflective beamforming at RIS & ------ \\
     \hline
\cite{luo2021reconfigurable}& Two-cell & Single-RIS & Maximizing ergodic weighted sum rate & OMA & Optimizing beamforming vector at BS and phase shift matrix at RIS & ------ \\
     \hline
\cite{rezaei2022energy}& Two-cell & Single-RIS & Maximizing energy efficiency & OMA & Optimizing antenna selection at BS and passive beamforming at RIS & ------ \\
     \hline
\cite{al2021performance}& Multi-cell & Multi-RIS & Maximizing system throughput & OMA & Proposed different combining scheme and root-based power assignment & ------ \\
     \hline
\cite{cai2021intelligent}& Multi-cell & Single-RIS & Minimizing total power consumption & OMA & Optimizing transmit beamforming at BS and reflective beamforming at RIS & ------ \\
     \hline
\cite{zhang2022meta}& Two-cell & Single-RIS & Maximizing sum rate & OMA & Optimizing beamforming at BS and RIS & ------ \\
     \hline     
\cite{9353406}& Multi-cell & Single-RIS & Maximizing the achievable sum rate & NOMA & Optimizing user, channel and power allocation along with phase shift & Perfect SIC \\
     \hline
\cite{ni2020intelligent}& Multi-cell & Single-RIS & Maximizing the achievable sum rate & NOMA & Optimizing user, channel and power allocation along with phase shift & Perfect SIC \\
     \hline
\cite{9681835}& Two-cell & Single-RIS & Maximizing sum rate & NOMA & Optimizing power, user clustering and phase shift design & Perfect SIC \\
     \hline   
\cite{zhang2021reconfigurable}& Two-cell & Two-RIS & Investigating outage performance & NOMA & Closed-form
analytical and asymptotic expressions for coverage probabilities & Perfect SIC \\
     \hline
\cite{xie2022star}& Multi-cell & Multi-RIS & Investigating coverage probability & NOMA & Closed-form solution for coverage probability and ergodic rate & Perfect SIC \\
     \hline
\cite{zhang2021multi}& Multi-cell & Multi-RIS & Investigating coverage probability & NOMA & Closed-form solution for coverage probability and ergodic rate & Perfect SIC \\
     \hline
[Our]& Multi-cell & Multi-RIS & Maximizing total spectral efficiency & NOMA & Optimizing BS power budget, NOMA users power allocation and RIS passive beamforming & Both perfect and imperfect SIC \\
     \hline
\end{tabular} 
\end{table*}

Researchers have recently studied the integration of NOMA with RIS in multi-cell wireless networks. For example, the research work in \cite{9353406,ni2020intelligent} proposed an efficient resource optimization framework to maximize the sum capacity of the multi-cell RIS-NOMA network. They optimized frequency allocation, power control, phase shift, and signal decoding using semidefinite relaxation and successive convex approximation. Another work in \cite{9681835} considered a two-cell RIS-NOMA network and jointly optimized power control, user clustering, and phase shift design to maximize the sum capacity of the system. Moreover, the authors of \cite{zhang2021reconfigurable} investigated the advantages of integration RIS with NOMA using stochastic geometry. The derived closed-form and asymptotic expressions for system coverage probability and ergodic capacity. The work in \cite{xie2022star} also proposed a stochastic geometry approach for multi-cell RIS-NOMA networks. The derived closed-form expressions for coverage probability and ergodic capacity. The work in \cite{zhang2021multi} exploited stochastic geometry to investigate the coverage performance of multi-cell RIS-NOMA wireless networks. They also modelled RIS-enabled channels consisting of direct and indirect transmission links.

\subsection{Motivation and Contributions}
Although extensive research works have been done on RIS-enabled wireless networks. However, the above recent works have the following shortcomings. 
\begin{enumerate}
    \item The works in \cite{buzzi2021ris, hashida2020intelligent, yu2022resource, cai2022irs, taghavi2022joint, pan2020multicell, huang2019reconfigurable,xu2022deep, qiu2022joint, xie2020max, luo2021reconfigurable, rezaei2022energy, al2021performance, cai2021intelligent, zhang2022meta} only consider OMA air interface technique in their proposed model, which has already been outperformed by NOMA technique.
    \item The works in \cite{9353406,ni2020intelligent,9681835} consider only a single RIS in two-cell and three-cell scenarios in their system models.
    \item The works of \cite{zhang2021reconfigurable,xie2022star,zhang2021multi} consider multiple RIS in their system; however, their focus is on performance analysis such as coverage probability and ergodic capacity, and they do not consider resource optimization.
    \item  The works which consider NOMA assume perfect signal decoding, which is challenging to obtain in such scenarios. Thus, considering the resource optimization problem for multi-cell RIS-NOMA system under imperfect signal decoding is not explored and is an open issue. A more detailed comparison with the existing works is also provided in Table 1.
\end{enumerate} 

\begin{figure*}[!t]
\centering
\includegraphics [width=0.70\textwidth]{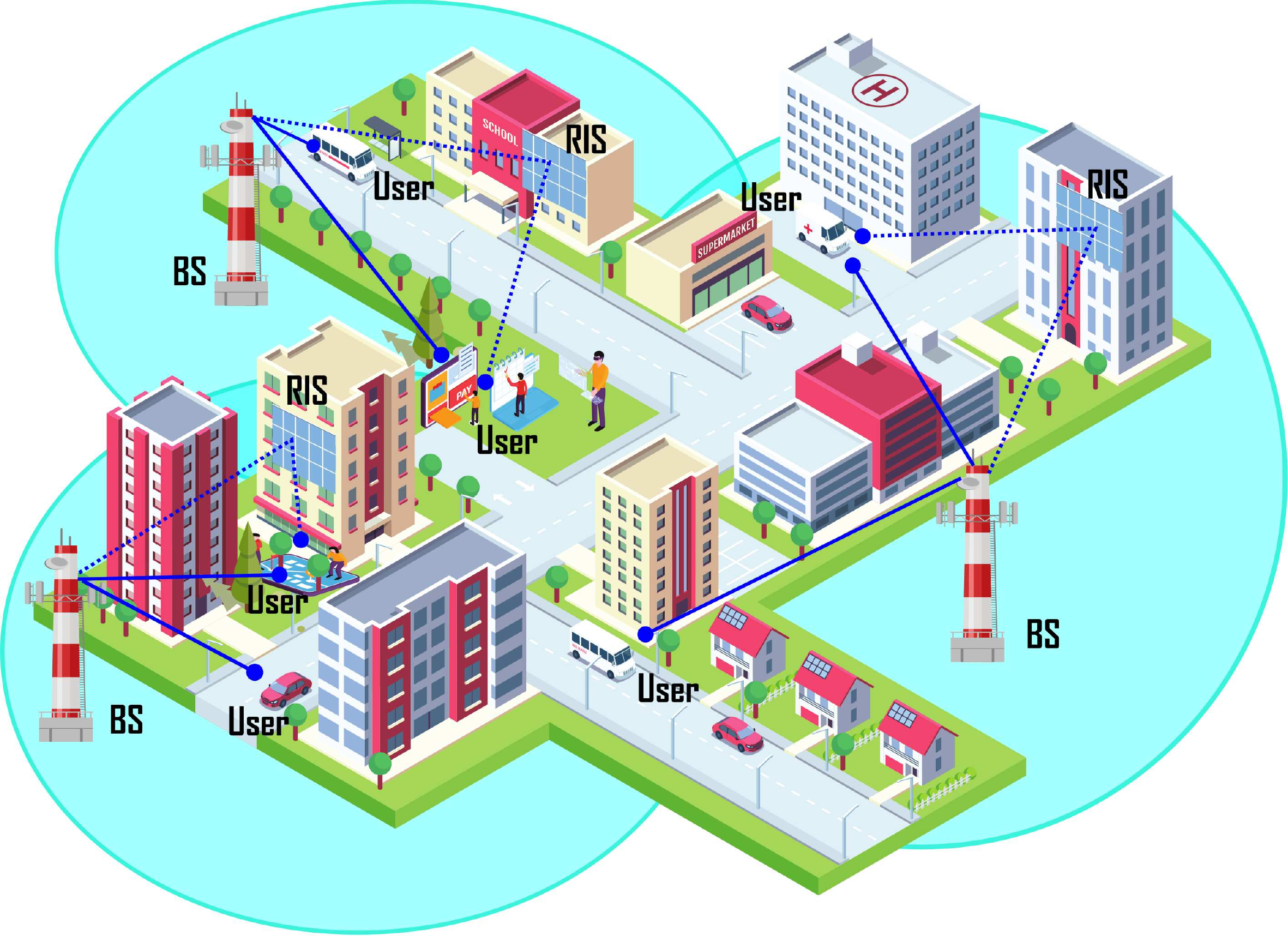}
\caption{System model of multi-cell RIS-NOMA network.}
\label{SM}
\end{figure*}
This paper considers a more realistic scenario to maximize the total achievable spectral efficiency of the system under signal decoding errors. In particular, our model consists of multi-cell ($>3$ cells) multi-RIS, which has not been investigated previously to the best of our knowledge. We simultaneously optimize the transmit power of the base station (BS), NOMA user power allocation, and passive beamforming of RIS in each cell while guaranteeing the minimum rate of each user. We exploit inner approximation and standard optimization methods for power allocation while successive convex approximation and Difference of Convex functions (DC) programming for the design of passive beamforming in each cell. Numerical results are presented to validate our proposed optimization scheme. The main contributions of this paper can also be summarized as follow.
\begin{itemize}
\item This paper considers a multi-cell RIS-NOMA model, where RIS in each cell assists the signal of NOMA users. In particular, in each cell, a BS in downlink transmits a superimposed signal to associated NOMA users. However, due to blockage of the signals and large-scale fading, a RIS is mounted on a strategic position to assist the signal delivery from BS to fading user. Therefore, the objective is to maximize the total achievable spectral efficiency of the system through optimal resource allocation. The proposed framework simultaneously optimizes the transmit power of BS, NOMA user power allocation and RIS phase shift design in each cell subject to the quality of services requirements.
\item The proposed optimization problem of sum rate maximization is formulated as non-convex due to inter-cell interference, imperfect signal decoding interference, and NOMA interference. Besides that, it is also coupled with multiple optimization variables, making it very complex to obtain an optimal solution. To make this problem more tractable, we decouple the original problem into sub-problems and transform it by inner approximation and successive convex approximation techniques. Next, we use a standard optimization method for
power allocation and DC programming, as well as standard semi-definite programming for passive beamforming design to obtain a sub-optimal yet efficient solution.
\item After obtaining an efficient solution, it is important to validate the proposed optimization scheme. To do so, we also provide numerical results based on Monte Carlo simulations. We provide results of three use case scenarios, i.e., proposed framework when considering perfect SIC decoding, proposed framework when RIS only assist signal for far NOMA user and proposed framework when RIS assist signal for both far and near users. Moreover, we check the impact of different simulation parameters on the system performance, such as the number of cells, minimum rate requirements, imperfect signal decoding errors, RIS elements, and transmit power of BS in each cell. The proposed optimization framework improves the system's spectral efficiency significantly and converges within a few iterations.
\end{itemize}
The remainder of this work is organized as follows. Section II explains the system model and states the channel models of the proposed framework. Section III formulates the total achievable spectral efficiency maximization problem and provides different steps of the proposed solution. Section IV presents and discusses the numerical results, while Section V concludes this paper with some attractive future research directions. 

\section{System and Channel Models}
As shown in Fig. 1, a multi-cell RIS-NOMA network is considered, where a BS in each cell serves two downlink users using the same spectrum resource. The set of BSs is denoted by $N$, where $\mathcal N=\{n|1, 2, 3, \dots,N\}$. We consider both users and BSs to be using a single-antenna scenario \cite{ahmed2022backscatter}, and the channel information is available in the system \cite{wei2021channel}. Considering an urban area, the signals of BSs can be blocked by various obstacles, i.e., buildings, trees, vehicles, and pedestrians. It results in very high fading and poor data rates. To enhance wireless coverage and improve the quality of services, a RIS system in each cell is considered, which consists of $K$ passive elements. We denote the diagonal reflection matrix of RIS as ${\bf\Theta}_n=\text{diag}\{\delta_{1}, \delta_{2}, \dots, \delta_{k}\}$, where $\delta_{k}$ is the reflection coefficient of $k^{th}$ element of RIS such that $\delta_{k}=\tilde{\beta_{k}}e^{j\theta_{k}}$, $ \theta_{k}\in[0,2\pi], \tilde{\beta_{k}}=1$, $\forall k$. The users served by BS $n$ are described as user $i$ and user $j$. Without sacrificing generality, we assume that both users $i$ and $j$ receive signals from BS via a direct link and from RIS via an indirect link. We also consider that user $i$ is close to the BS $n$ and has a high channel gain, whereas user $j$ is far from the BS $n$ and has a low channel gain. As a result, the direct signal from BS $n$ to user $i$ is stronger than the signal from the RIS-assisted link. On the other side, the direct signal from user $j$ to BS $n$ is weaker than its RIS-assisted link. Following the above observation, the direct channel of users $i$ and $j$ from BS $n$ can be sorted as $g_{i,n}\geq g_{j,n}$. Here, $g_{\iota,n}=\bar{g}_{\iota,n}D^{-\eta/2}_{\iota,n}$ denotes the channel gain of the direct link between BS $n$ and user $\iota\in\{i,j\}$ such that $\bar{g}_{\iota,n}\in\mathcal{CN}(0,1)$ is the Rayleigh fading coefficient and $D_{\iota,n}$ shows the distance between user and BS while $\eta$ is the path-loss exponent \cite{khan2022energy}\footnote{Unless mention otherwise, the channels in the system are modelled similarly but the details are omitted here for simplicity.}. If $x_n$ is the superimposed signal, then the transmit signal of BS $n$ can be represented as:
\begin{align}
x_n=\sqrt{P_n\alpha_{i,n}}x_{i,n}+\sqrt{P_n\alpha_{j,n}}x_{j,n},
\end{align}
where $P_n$ denotes the transmit power of BS $n$, $\alpha_{i,n}$ and $\alpha_{j,n}$ are the power allocation coefficients while $x_{i,n}$ and $x_{j,n}$ represent the unit power signals, respectively. The received signal that user $i$ receives from BS $n$ through both direct and RIS assisted links can be written as:
\begin{align}
y_{i,n}&=\underbrace{(g_{i,n}+\textbf{h}^H_{i,k,n}{\bf\Theta}_n\textbf{f}_{i,k,n})x_{n}}_{\text{desired superimposed signal}}+\underbrace{\omega_{i,n}}_{\text{AWGN}}\nonumber\\
&+\underbrace{\sum\limits_{n'=1,n'\neq n}^N(g_{i,n'}+\textbf{h}^H_{i,k,n'}{\bf\Theta}_n\textbf{f}_{i,k,n'})x_{n'}}_{\text{co-channel interference from neighboring cell}},\label{1}
\end{align}
 where $\omega_{i,n}\sim CN(0,\sigma^2)$ is the additive white Gaussian noise (AWGN) of user $i$ with $\sigma^2$ variance. Moreover, $\textbf{h}_{i,k,n}\in{K\times1}$ denotes Rician fading channels from BS $n$ to RIS while $\textbf{f}_{i,k,n}\in K\times 1$ represents Rayleigh fading channels from RIS to user $i$. Accordingly, the received signal that user $j$ receives through RIS assisted link can be expressed as:
\begin{align}
y_{j,n}&=\underbrace{(g_{j,n}+\textbf{h}^H_{j,k,n}{\bf\Theta}_n\textbf{f}_{j,k,n})x_{n}}_{\text{desired superimposed signal}}+\underbrace{\omega_{j,n}}_{\text{AWGN}}\nonumber\\
&+\underbrace{\sum\limits_{n'=1,n'\neq n}^N(g_{j,n'}+\textbf{h}^H_{j,k,n'}{\bf\Theta}_n\textbf{f}_{j,k,n'})x_{n'}}_{\text{co-channel interference from neighboring cell}},\label{2}
\end{align}
where $\omega_{j,n}\sim CN(0,\sigma^2)$ is the additive white Gaussian noise (AWGN) of user $j$ with $\sigma^2$ variance. Moreover, $\textbf{h}_{j,k,n}\in{K\times1}$ denotes Rician fading channels from BS $n$ to RIS $m$ while $\textbf{f}_{j,k,n}\in K\times 1$ represents Rayleigh fading channels from RIS to user $j$, respectively.

Further, let  $\boldsymbol{\delta}_n=[\delta_1,\delta_2,\ldots, \delta_k]^T \in \mathbb{C}^{K\times 1}$ denoted the vector of diagonal elements of passive beamforming matrix $\bf{\Theta}_n$. Then, we can rewrite the Eqs. (\ref{1}) and (\ref{2}) as follows  
\begin{align}
	y_{i,n}&=\underbrace{(g_{i,n}+{\boldsymbol{\delta}_n^H}\text{diag}(\textbf{h}_{i,k,n}) \textbf{f}_{i,k,n})x_{n}}_{\text{desired superimposed signal}}\nonumber\\
	&+\underbrace{\sum\limits_{n'=1,n'\neq n}^N (g_{i,n'}+{\boldsymbol{\delta}_n^H}\text{diag}(\textbf{h}_{i,k,n'}) \textbf{f}_{i,k,n'})x_{n'}}_{\text{co-channel interference from neighboring cell}},\label{1a}
\end{align}

\begin{align}
	y_{j,n}&=\underbrace{(g_{j,n}+{\boldsymbol{\delta}_n^H}\text{diag}(\textbf{h}_{j,k,n}) \textbf{f}_{j,k,n})x_{n}}_{\text{desired superimposed signal}}\nonumber\\
	&+\underbrace{\sum\limits_{n'=1,n'\neq n}^N (g_{j,n'}+({\boldsymbol{\delta}_n^H}\text{diag}(\textbf{h}_{j,k,n'}) \textbf{f}_{j,k,n'})x_{n'}}_{\text{co-channel interference from neighboring cell}},\label{2a}
\end{align}

Following the NOMA protocol, we assume that the user with strong channel conditions is capable of applying SIC to subtract the signal of the weak user before decoding its own signal. Based on (\ref{1a}) and (\ref{2a}), the signal-to-interference plus noise ratio (SINR) of user $i$ and user $j$ can be computed as:
\begin{align}
\gamma_{i,n}&=\frac{P_n\alpha_{i,n}(|g_{i,n}+{\boldsymbol{\delta}_n^H}\text{diag}(\textbf{h}_{i,k,n}) \textbf{f}_{i,k,n}|^2)}{P_n\alpha_{j,n}(|g_{i,n}+{\boldsymbol{\delta}_n^H}\text{diag}(\textbf{h}_{i,k,n}) \textbf{f}_{i,k,n}|^2)\beta+I_{i,n'}+\sigma^2}, 
\end{align}
\begin{align}
\gamma_{j,n}&=\frac{P_n\alpha_{j,n}|g_{j,n}+{\boldsymbol{\delta}_n^H}\text{diag}(\textbf{h}_{j,k,n}) \textbf{f}_{j,k,n}|^2}{P_n\alpha_{i,n}|g_{j,n}+{\boldsymbol{\delta}_n^H}\text{diag}(\textbf{h}_{j,k,n}) \textbf{f}_{j,k,n}|^2+I_{j,n'}+\sigma^2},
\end{align}
where $I_{i,n'}$ and  $I_{j,n'}$ denotes the co-channel interference at user $i$ and user $j$, respectively, and $\beta$ denotes the parameter of SIC decoding error, respectively. 
Thus, the corresponding data rate of $\gamma_{i,n}$ and $\gamma_{j,n}$ can be computed as follows
\begin{align}
	R_{i,n} = \log_{2}(1+\gamma_{i,n}), \label{1111}
\end{align}
\begin{align}
	R_{j,n} = \log_{2}(1+\gamma_{j,n}), \label{2222}
\end{align}
Finally, the sum-rate of the considered system can be calculated as
\begin{align}
	R = \sum\limits_{n=1}^N (R_{i,n}+ R_{j,n}), \label{333}
\end{align}

\section{Problem Formulation and Proposed Solution}
This work seeks to maximize the spectral efficiency of the multi-cell RIS-NOMA network by optimizing the transmit power of BS and the reflection matrix of RIS in each cell. The optimization problem can be formulated as:
\allowdisplaybreaks
\begin{alignat}{2}
\underset{{\boldsymbol{\alpha}_n, P_n, {\boldsymbol{\delta}_n}}}{\text{maximize}} & \sum\limits_{n=1}^N\{R_{i,n}+ R_{j,n}\}\label{5}\\
 s.t.\ T_1:\ & R_{i,n}\geq R_{min},\forall i,n, \nonumber\\
 T_2:\ &R_{j,n}\geq R_{min},\forall j,n,  \nonumber \\
 T_3:\ & 0\leq P_n\leq P_{tot}, \forall n,  \nonumber\\
 T_4:\ & \theta_{k,n}\in[0,2\pi], \forall k, n,\nonumber\\
 T_5:\ &0\leq \alpha_{i,n}\leq 1,0\leq \alpha_{j,n}\leq 1, \forall i,j,n,\nonumber\\
 T_6:\ &\alpha_{i,n}+\alpha_{j,n}=1, \forall i,j,n, \nonumber
\end{alignat}
where $\boldsymbol{\alpha_n}=\{\alpha_{i,n},\alpha_{j,n}\}$. The constraints $T_1$ and $T_2$ ensure the minimum data rate of user $i$ and user $j$. Constraint $T_3$ controls the transmit power of BS in each cell. Constraints $T_4$ is invoked for phase shift and reflection coefficient at RIS. Constraints $T_5$ and $T_6$ are the power distribution restrictions of the NOMA systems. Note that problem (\ref{5}) is non-convex because of the objective function and the constraints $T_1$ and $T_2$. Moreover, the problem (\ref{5}) is also coupled on multiple optimization variables. Therefore, obtaining an optimal global solution is very challenging and complex. To overcome this, we decouple the original problem \eqref{5} into sub-problems, transform them and then solve them iteratively.

\subsection{Power-allocation Optimization}
For any given reflection matrix at RIS, the optimization problem defined in (\ref{5}) can be simplified to a power-allocation problem as follows
\begin{alignat}{2}
\underset{{\boldsymbol{\alpha}_n, P_n}}{\text{maximize}} & \sum\limits_{n=1}^N\{R_{i,n}+R_{j,n}\}\label{s5}\\
 s.t.\quad & T_1, T_2, T_3, T_5, T_6,\nonumber
\end{alignat}
where (\ref{s5}) is the power allocation problem at BS in each cell. This problem is still non-convex. For the SINR of user $i$ and user $j$ associated with BS $n$, we can efficiently replace $a_{j,n}$ by $1-a_{i,n}$. Substituting $a_{j,n}=1-a_{i,n}$ in the problem removes the need for $T_6$ and also makes the problem more tractable. Next, we introduce two arbitrary positive variables $\psi_{i,n}$ and $\psi_{j,n}$ \cite{8886077} such that the problem (\ref{5}) for any given reflection matrix at RIS can be effectively reformulated as:
\begin{alignat}{2}
&\underset{{\boldsymbol{\{\psi_{i,n},\psi_{j,n}\}\{\alpha}_n, P_n\}}}{\text{maximize}} \sum\limits_{n=1}^N\{\psi_{i,n}+\psi_{j,n}\}\label{13}\\
 s.t.\ & T'_1:\ R_{i,n}\geq \psi_{i,n},\forall i,n, \nonumber\\
& T'_2:\ R_{j,n}\geq \psi_{j,n},\forall j,n,  \nonumber \\
& \ T_3, T_5, T_6, \nonumber
\end{alignat}
Next, to further simplify, we define $\chi_{i,n}=|g_{i,n}+{\boldsymbol{\delta}_n^H} \text{diag}(\textbf{h}_{i,k,n}) \textbf{f}_{i,k,n}|^2$ and $\chi_{j,n}=|g_{j,n}+{\boldsymbol{\delta}_n^H} \text{diag}(\textbf{h}_{j,k,n}) \textbf{f}_{j,k,n}|^2$ and transform the problem \eqref{13} to its equivalent problem (please refer to Appendix A) as follows:
\begin{alignat}{2}
&\underset{{\boldsymbol{\{\small{\psi_{i,n},\psi_{j,n}}\}\{\alpha}_n, P_n\}}}{\text{maximize}} \sum\limits_{n=1}^N\{\psi_{i,n}+\psi_{j,n}\}\label{7}\\ 
 s.t.\ &T'_1: \log\left({P_n(1-\alpha_{i,n})\chi_{i,n}\beta+I_{i,n'}+\sigma^2+P_n\alpha_{i,n}\chi_{i,n}} \right)\nonumber\\ &\geq\psi_{i,n}\log(2)+\log(P_n(1-\alpha_{i,n})\chi_{i,n}\beta+I_{i,n'}+\sigma^2), \forall i,n, \nonumber\\
 &T'_2:\log\left({P_n\alpha_{i,n}\chi_{j,n}+I_{j,n'}+\sigma^2+P_n(1-\alpha_{i,n})\chi_{j,n}} \right)\nonumber\\&\geq\psi_{j,n}\log(2)+\log(P_n\alpha_{i,n}\chi_{j,n}+I_{j,n'}+\sigma^2), \forall j,n, \nonumber\\
& \ T_3, T_5, T_6, \nonumber
\end{alignat} 
where it can be observed that problem \eqref{7} is still non-convex due to the logarithm function in the second term of constraint $T'_1$ and constraint $T'_2$. To address this issue, we add two auxiliary variables $\amalg_{i,n}$ and $\amalg_{j,n}$ such that the problem \eqref{7} can be efficiently reformulated  as:
\begin{alignat}{2}
&\underset{{\boldsymbol{\{\small{\psi_{i,n},\psi_{j,n},\amalg_{i,n},\amalg_{j,n}}\}\{\alpha}_n, P_n\}}}{\text{maximize}} \sum\limits_{n=1}^N\{\psi_{i,n}+\psi_{j,n}\}\label{8}\\ 
 s.t.\ T'_1:& \log (P_n(1-\alpha_{i,n})\chi_{i,n}\beta+I_{i,n'}+\sigma^2+P_n\alpha_{i,n}\chi_{i,n})\nonumber\\&\geq \psi_{i,n}\log(2)+\amalg_{i,n},\nonumber\\
 T'_2:&\log(P_n(1-\alpha_{i,n})\chi_{j,n})\nonumber\\&\geq\psi_{j,n}\log(2)+\amalg_{j,n},\nonumber\\
T'_{1,a}:& P_n(1-\alpha_{i,n})\chi_{i,n}\beta+I_{i,n'}+\sigma^2\le e^{\amalg_{i,n}},\nonumber\\
T'_{2,a}:&P_n\alpha_{i,n}\chi_{j,n}+I_{j,n'}+\sigma^2\le e^{\amalg_{j,n}},\nonumber\\
& \ T_3, T_5, T_6, \nonumber
\end{alignat} 
where we can observe that constraints $T'_1$ and $T'_2$ are transformed into convex expressions. However, the addition of new constraints $T'_{1,a}$ and $T'_{2,a}$ make the problem \eqref{8} difficult to solve and find the optimal value. Therefore, due to the convex function $e^x$ in $T'_{1,a}$ and $T'_{1,b}$, we can use the inner approximation method, which uses the first-order approximation of the exponential function on the right side of the constraint $T'_{1,a}$ and $T'_{2,a}$. The approximated problem can be written below:
\begin{alignat}{2}
&\underset{{\boldsymbol{\{\small{\psi_{i,n},\psi_{j,n},\amalg_{i,n},\amalg_{j,n}\}\{\psi}_n, P_n}\}}}{\text{maximize}} \sum\limits_{n=1}^N\{\psi_{i,n}+\psi_{j,n}\}\label{16}\\
 s.t.\quad & T'_{1,b}: P_n(1-\alpha_{i,n})\chi_{i,n}\beta+I_{i,n'}+\sigma^2\nonumber\\
& \le e^{\amalg_{i,n}^o}\left(\amalg_{i,n}-\amalg_{i,n}^o+1\right),\nonumber\\
& T'_{2,b}: P_n\alpha_{i,n}\chi_{j,n}+I_{j,n'}+\sigma^2\nonumber\\
&\le e^{\amalg_{j,n}^o}\left(\amalg_{i,n}-\amalg_{i,n}^o+1\right),\nonumber\\
& T'_1,T'_2, T_3, T_5, T_6, \nonumber
\end{alignat} 
According to \eqref{16}, the objective function and constraints are convex for the given values of $\amalg_{i,n}^o$ and $\amalg_{j,n}^o$ and can be efficiently solved using a standard solver, such as CVX.

\subsection{Passive-beamforming Optimization}
In this subsection, we compute the passive-beamforming vector $\boldsymbol{\delta}_n$ of RIS for the given values of $\alpha^*_n$ and $P^*_n$. the problem (\ref{5}) can recast as:
\begin{alignat}{2}
\underset{{\boldsymbol{\delta}_n}}{\text{maximize}} & \sum\limits_{n=1}^N (R_{i,n}+ R_{j,n})\label{ss52}\\
 s.t.\quad  & T'_{1,c}:  R_{i,n}\geq R_{min},\forall n, \nonumber\\
 & T'_{2,c}:  R_{j,n}\geq R_{min},\forall n, \nonumber\\
  & T'_{3,c}:  \theta_{k,n}\in[0,2\pi], \forall k,n, \nonumber
\end{alignat}
Next, to compute the solution of $\boldsymbol{\delta}_n$, the main steps are summarized as follows:

\noindent \textbf{Step 1:} Since the objective function of the optimization problem in \eqref{ss52} is highly non-convex, it is very hard to be tackled. Therefore, we first adopt successive convex approximation (SCA) \cite{5165179} to reduce the complexity of the sum-rate in \eqref{ss52}. By exploiting the SCA technique, the achievable rates of NONA users, $R_{i,n}$, $R_{j,n}$, can be updated as follows 
\begin{align}
	\bar{R}_{i,n} = \zeta_{i,n}\log_2(\gamma_{i,n})+\Omega_{i,n}, \label{11111}
\end{align}
\begin{align}
	\bar{R}_{j,n} = \zeta_{j,n}\log_2(\gamma_{j,n})+\Omega_{j,n}, \label{22222}
\end{align}
where $\zeta_{e,n}=\dfrac{\gamma_{e,n}}{1+\gamma_{e,n}}, \forall e=\{i,j\},$ and $\Omega_{e,n}=\log_{2}(1+\gamma_{e,n})-\dfrac{\gamma_{e,n}}{1+\gamma_{e,n}}\log_{2}(\gamma_{e,n}),\forall e=\{i,j\}$. Hence, based on Eqs. \eqref{11111} and \eqref{22222}, the sum-rate of the considered system can be updated as follows   
\begin{align}
	\bar{R} = \sum\limits_{n=1}^N (\bar{R}_{i,n}+ \bar{R}_{j,n}), \label{3333}
\end{align}
Hence, the optimization problem defined by \eqref{ss52} can be recast as follows 
\begin{alignat}{2}
	\underset{{\boldsymbol{\delta}_n}}{\text{maximize}} & \sum\limits_{n=1}^N (\bar{R}_{i,n}+ \bar{R}_{j,n}) \label{sss52}\\
	s.t.\quad  & T'_{1,d}:  R_{i,n}\geq R_{min},\forall n, \nonumber\\
	& T'_{2,d}:  R_{j,n}\geq R_{min},\forall n, \nonumber\\
	& T'_{3,d}:  \theta_{k,n}\in[0,2\pi], \forall k,n, \nonumber
\end{alignat}
\noindent \textbf{Step 2:} Let, $\textbf{W}_{i,n}=diag(\textbf{h}_{i,k,n})\textbf{f}_{i,k,n}$, $\textbf{W}_{j,n}=diag(\textbf{h}_{n,k,n}) \textbf{f}_{j,k,n}$,  $\boldsymbol{\bar{{\delta}}}_n=\begin{bmatrix} \boldsymbol{\delta}_n\\ 1\end{bmatrix}$. Also, define a matrix  $\mathbf{A}=\boldsymbol{\bar{{\delta}}}_n \boldsymbol{\bar{{\delta}}}^H_n$, where $\mathbf{A} \succeq 0$, $rank (\mathbf{A}) = 1$. Then, the achievable rates of NONA users can be updated as  
\begin{align}
	\bar{R}^*_{i,n} &=\Big[\zeta_{i,n}\Big\{\log_{2}\Big(P_n\alpha_{i,n} \times \Big(Tr.(\mathbf{G}_{i,n}\mathbf{A})+|g_{i,n}|^2\Big)\Big)\nonumber\\
	&-\log_{2}\Big(P_n\alpha_{j,n}\Big(Tr.(\mathbf{G}_{i,n}\mathbf{A})+|g_{i,n}|^2\Big)+\bar{\sigma}^{2}_{i}\Big)\beta\Big\} +\Omega_{i,n}\Big], \label{111111}
\end{align} 
and,
\begin{align}
	\bar{R}^*_{j,n} &=\Big[ \zeta_{j,n}\Big\{\log_{2}\Big(P_n\alpha_{j,n} \times \Big(Tr.(\mathbf{G}_{j,n}\mathbf{A})+|g_{j,n}|^2\Big)\Big)\nonumber\\
	&-\log_{2}\Big(P_n\alpha_{i,n} \Big(Tr.(\mathbf{G}_{j,n}\mathbf{A})+|g_{j,n}|^2\Big)+\bar{\sigma}^{2}_{j}\Big)\Big\} +\Omega_{j,n}\Big], \label{222222}
\end{align}
where $\bar{\sigma}^{2}_{i}=I_{i,n'}+\sigma^2$, $\bar{\sigma}^{2}_{j}=I_{j,n'}+\sigma^2$, $Tr.(.)$ denotes the trace property, and
\begin{align}
	\mathbf{G}_{i,n}=\begin{bmatrix}
		\textbf{W}_{i,n} \textbf{W}_{i,n}^H	& \textbf{W}_{i,n} g_{i,n}^H\\ 
		\textbf{W}_{i,n}^H g_{i,n}& 0
	\end{bmatrix},\label{1111111}
\end{align}
\begin{align}
			\mathbf{G}_{j,n}=\begin{bmatrix}
			\textbf{W}_{j,n} \textbf{W}_{j,n}^H	& \textbf{W}_{j,n} g_{j,n}^H\\ 
			\textbf{W}_{j,n}^H g_{j,n}& 0
	\end{bmatrix},\label{2222222}
\end{align}
Hence, the sum-rate of the considered system can be updated as follows
\begin{align}
	\bar{R}^* = \sum\limits_{n=1}^N (\bar{R}^*_{i,n}+ \bar{R}^*_{j,n}), \label{33333}
\end{align}
\noindent \textbf{Step 3:}
It is important to note that $\bar{R}^*_{j,n}$ in \eqref{222222} is the difference of two concave functions. Thus, the considered optimization problem defined by Eq. \eqref{sss52} is non-convex. Then, for $f_{j,1}(\mathbf{A})=\log_{2}\Big(P_n\alpha_{j,n} \times \Big(Tr.(\mathbf{G}_{j,n}\mathbf{A})+|g_{j,n}|^2\Big)$ and $f_{j,2}(\mathbf{A})=\log_{2}\Big(P_n\alpha_{i,n} \Big(Tr.(\mathbf{G}_{j,n}\mathbf{A})+|g_{j,n}|^2\Big)+\bar{\sigma}^{2}_{j}\Big)$, we can write $\bar{R}^*_{j,n}$ in \eqref{222222} as a function of $A$ given as follows
\begin{align}
	&\overline{R_{j,n}^*}=\zeta_{j,n}[f_{j,1}(\mathbf{A})-f_{j,2}(\mathbf{A})]+\Omega_{j,n},
\end{align}

Further, by exploiting a low complexity technique, known as the difference of convex (DC) programming,  we replace $f_{j,2}(\mathbf{A})$ with its first-order Taylor approximation as: 
\begin{align}
	f_{j,2}(\mathbf{A})&\leq f_{j,2}(\mathbf{A}^{(i)})+Tr.((f'_{j,2}(\mathbf{A}^{(i)}))^{H}(\mathbf{A}-\mathbf{A}^{(i)} ))  \nonumber\\
	& \triangleq \overline{f_{j,2}(\mathbf{A})}, \label{333333}
\end{align}
where $\mathbf{\mathbf{A}}^{(i)}$ denotes the value of $\mathbf{A}$ obtained in the $i^{th}$ iteration. Then, $\overline{R_{j,n}^*}$ can be written as  
\begin{align}
	&\overline{R_{j,n}^{**}}=\zeta_{j,n}[f_{j,1}(\mathbf{A})- \overline{f_{j,2}(\mathbf{A})}]+\Omega_{j,n},
\end{align}
Hence, the sum-rate of the considered system defined in \eqref{33333} can be updated as follows
\begin{align}
	\tilde{R} = \sum\limits_{n=1}^N (\bar{R}^*_{i,n}+ \overline{R_{j,n}^{**}}), \label{3333333}
\end{align}
Consequently, the optimization problem defined in Eq \eqref{sss52} can be reformulated as 
\begin{subequations}\label{Prob:EE_88}
	\begin{align}
	\underset{\mathbf{A}}{\text{maximize}} & \sum\limits_{n=1}^N (\bar{R}^*_{i,n}+ \overline{R_{j,n}^{**}}) \label{ssss52}\\
	s.t.\quad  &  \bar{R}^*_{i,n}\geq R_{min},\forall n, \\
	&   \overline{R_{j,n}^{**}}\geq R_{min},\forall n,\\
    & \mathbf{A}_{k,k} \leq 1, \forall k,\\
    &\mathbf{A} \succeq 0,\\
    & rank (\mathbf{A}) = 1. 
	\end{align}
\end{subequations}
\noindent \textbf{Step 4:} It is important to note that the above optimization problem in \eqref{Prob:EE_88} is still non-convex in nature due to non-convex Rank-1 constraint defined in Eq. (31f). Since $\mathbf{A} \in \mathbb{C}^{K\times K}$, Tr.($\mathbf{A})>0$, is positive semi-definite (PSD) matrix, we exploit DC programming technique to approximate the Rank-1 constraint given in Eq. (31f) as follows: 
\begin{equation}
rank (\mathbf{A}) = 1 \Leftrightarrow Tr.(\mathbf{A}) -||\mathbf{A}||_2 =0, \label{737}
\end{equation}
where $Tr.(\mathbf{A}) = \sum\limits_{k=1}^{K} \nu_k(\mathbf{A})$ and $||\mathbf{A}||_2=\nu_1(\mathbf{A})$ represent the trace and spectral-norm of $\mathbf{A}$. $\nu_k$ is the $k^{th}$ largest singular value of $\mathbf{A}$. Please note that the Rank-1 constraint in \eqref{737} is still non-convex. Therefore, we exploit the Taylor expansion to approximate the spectral-norm of $\mathbf{A}$ given as follows 
\begin{align}
	||\mathbf{A}||_2 &\geq ||\mathbf{A}^{(i)}||_2 + Tr.\Big(\mathcal{J}^{(i)}_{max}(\mathbf{A}^{(i)}) {\mathcal{J}^{(i)}_{max}(\mathbf{A}^{(i)})}^H(\mathbf{A}-\mathbf{A}^{(i)})\Big)\nonumber , \\
	&  \triangleq \overline{||\mathbf{A}||_2},\label{744}
\end{align}
where $\mathcal{J}^{(i)}(\mathbf{A}^{(m)})$ represents the eigenvector corresponding to the largest singular value of $\mathbf{A}$.

\noindent \textbf{Step 5:} Finally, we add the above approximated Rank-1 constraint in the objective function of the optimization problem, defined in \eqref{Prob:EE_88}, as a penalty term given as follows\\\\
\begin{subequations}\label{Prob:EE_888}
	\begin{align}
		\underset{\mathbf{A}}{\text{maximize}} & \sum\limits_{n=1}^N (\bar{R}^*_{i,n}+ \overline{R_{j,n}^{**}})-\Xi\Big(Tr.(\mathbf{A})-\overline{||\mathbf{A}||_2}\Big) \label{sssss52}\\
		s.t.\quad  &  \bar{R}^*_{i,n}\geq R_{min},\forall n, \\
		&   \overline{R_{j,n}^{**}}\geq R_{min},\forall n,\\
		& \mathbf{A}_{k,k} \leq 1, \forall k,\\
		&\mathbf{A} \succeq 0,
	\end{align}
\end{subequations}
where $\Xi>>0$ denotes the penalty factor for the above approximated Rank-1 constraint. Consequently, the above optimization problem defined in \eqref{Prob:EE_888} is in the form of a standard semi-definite programming (SDP) problem, where the solution of this SDP problem can be easily obtained using MOSEK-enabled CVX toolbox of MATLAB \cite{Jzhou}.

\subsection{Complexity Analysis}
In this work, we have proposed a new optimization framework for the multi-cell RIS-NOMA networks. The proposed spectral efficiency problem has been solved in two-steps, i.e., power allocation and passive beamforming. In the first-step, the transmit power of BS $P_n$ and power allocation coefficient of NOMA users $\alpha_{i,n}, \alpha_{j,n}$ in each cell has been calculated for any given passive beamforming $\delta_n$ of RIS. After computing these values, we substitute it in the proposed problem and then calculate the passive beamforming of RIS in second-step. The detailed solution process is also given in {\bf Algorithm 1}. It is important to discuss the complexity of the proposed multi-cell RIS-NOMA optimization scheme. Here, the term complexity refers to the number of iterations required for the convergence of optimization variables. The complexity of the first-step for calculating the transmit power in a given iteration can be expressed as $\mathcal O(2N)$. Accordingly, the complexity of the proposed algorithm in second-step for computing passive beamforming is $\mathcal O(NK^{3.5})$ \cite{wright1997primal}. If $T$ is the required number of iterations in the proposed optimization process, then the overall complexity of the proposed {\bf Algorithm 1} becomes $\mathcal O\{T(2N+NK^{3.5})\}$.

\begin{algorithm}[!t]\small
{\bf Initialization:} Define all parameters of the system and $\Phi^o$=\{$\amalg_{i,n}^o$ , $\amalg_{j,n}^o$, $\psi_{i,n}^{o}$, $\psi_{j,n}^{o}$\}, \textbf{error}.
 
 {\bf Step 1:} Compute $\alpha_{i,n}$, $\alpha_{j,n}$, and $P_n$ for the given ${\boldsymbol{\delta}_n}$
         
    \While{\textbf{error}$\le \epsilon$ }{
        solve \eqref{16} to find the value of $\Phi^*$ = \{$\amalg_{i,n}^*$, $\amalg_{j,n}^*$, $\psi_{i,n}^{*}$, $\psi_{j,n}^{*}$ $\alpha_{i,n}^*$, $\alpha_{j,n}^*$, and $P_n^*$\} \\
        // Given the value of all the decision variables, calculate the error\\
        \textbf{error}= $\Phi^o$-$\Phi^*$
        }  
{\bf Step 2:} Now with $\alpha^*_{i,n},\alpha^*_{j,n},P^*_n$, we compute ${\boldsymbol{\delta}_n}$
    
    \For{$n=1:N$}{Find ${\boldsymbol{\delta}_n}$ by solving the standard semi-definite programming in \eqref{Prob:EE_888} using MOSEK-enabled CVX toolbox of MATLAB.
    }

    Return $\alpha^*_{i,n}$, $\alpha^*_{j,n}$, $P^*_n$,  ${\boldsymbol{\delta}^*_n}$
    \caption{Proposed optimization Scheme}
   \end{algorithm}  

\section{Numerical Results}
In this section, we present the numerical results of the proposed optimization scheme. Unless mentioned otherwise, the simulation parameters are set as follows: The average channels are obtained from 10000 realizations; the power budget of each BS is 20 dBm; the number of passive elements on each RIS is 50; consider full frequency reuse such that each BS receives co-channel interference from neighbouring BSs; the number of users associated with each BS is 2; the path loss exponent is 3; the number of BSs is 10; and the imperfect SIC decoding error $\beta=0.1$. Moreover, the channels from BS to RIS undergo Rician fading, while the channels from BS to users and from RIS to users undergo Rayleigh fading. Furthermore, the variance of the AWGN is set as $\sigma^2=0.1$. We provide and compare the results of the following three proposed frameworks under the same optimization variables, i.e., power allocation of BS and users, as well as the passive beamforming of RIS in each cell.
\begin{enumerate}
    \item Proposed framework 1: This is the communication scenario proposed in Section III, where in each cell, users receive the signal from BS through both direct link and RIS-assisted link. This communication scenario considers perfect SIC decoding on the user's side.
    \item Proposed framework 2: This is the communication scenario, where in each cell, the weak user receives the signal from BS through both direct and RIS-assisted links while the strong user receives the signal from BS through only direct link under perfect SIC decoding.
    \item Proposed framework 3: It refers to the communication scenario, where users in each cell receive the signal from BS through both direct and RIS-assisted links under the assumption of imperfect SIC decoding error.
\end{enumerate}

\begin{figure}[!t]
\centering
\includegraphics[width=0.5\textwidth]{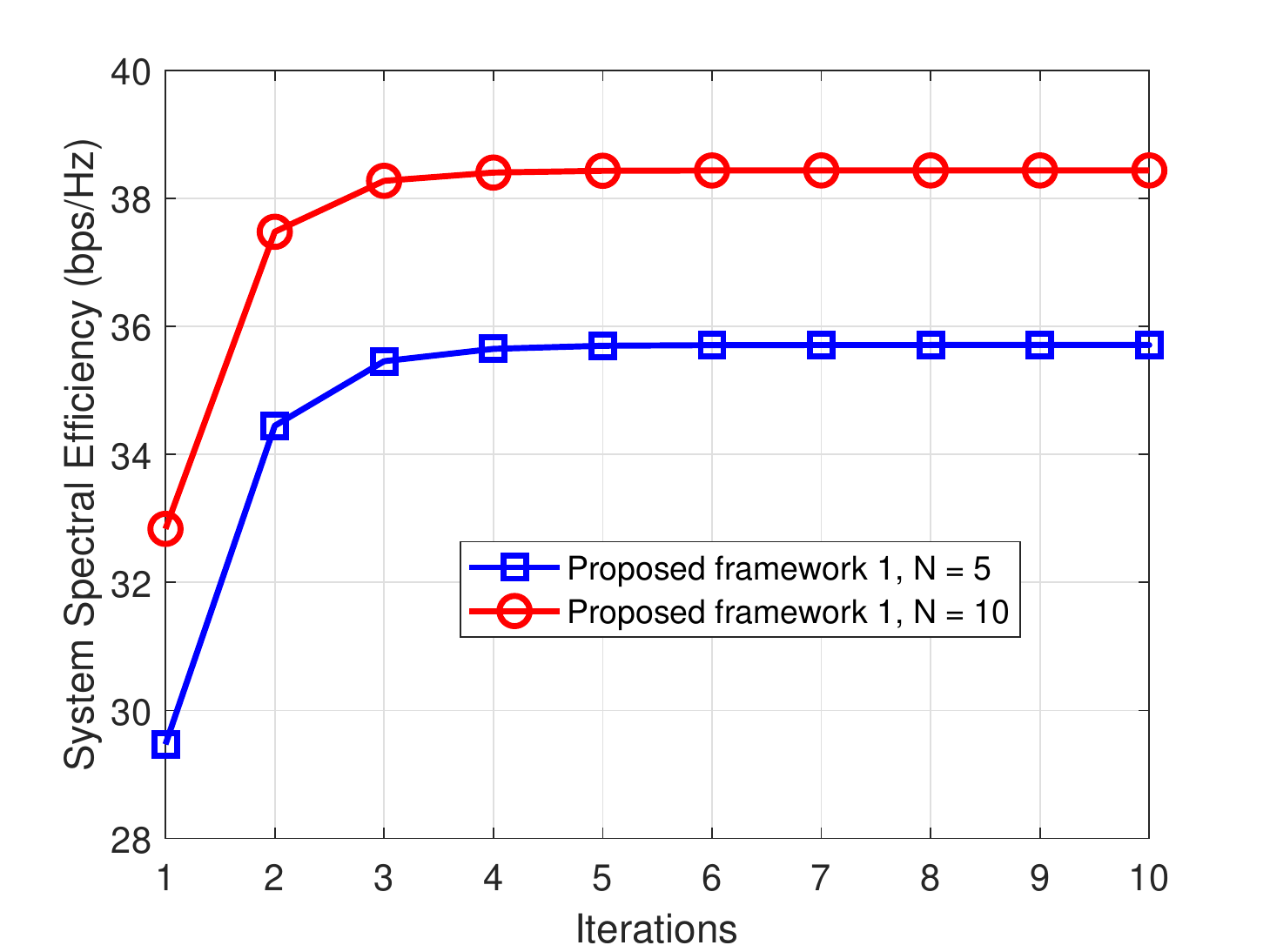}
\caption{Convergence of system spectral efficiency for the proposed framework 1.}
\label{Fig1}
\end{figure}
First, it is important to check the convergence behaviour of the proposed optimization scheme. Therefore, we plot the system's spectral efficiency versus the number of iterations, as shown in \ref{Fig1}. It can be witnessed from the figure that the system converges within a few iterations, and increasing the number of interfering cells has a negligible impact on the convergence time. For example, when $N = 5$ and $N = 10$, the proposed framework 1 converges when the number of iterations exceeds 4. It indicates that the proposed optimization scheme performs significantly high with very low complexity. It is worth mentioning here that the other communication scenarios, namely, proposed framework 2 and proposed framework 3 have similar complexity due to the same optimization variables and solution approach.

\begin{figure}[!t]
\centering
\includegraphics[width=0.5\textwidth]{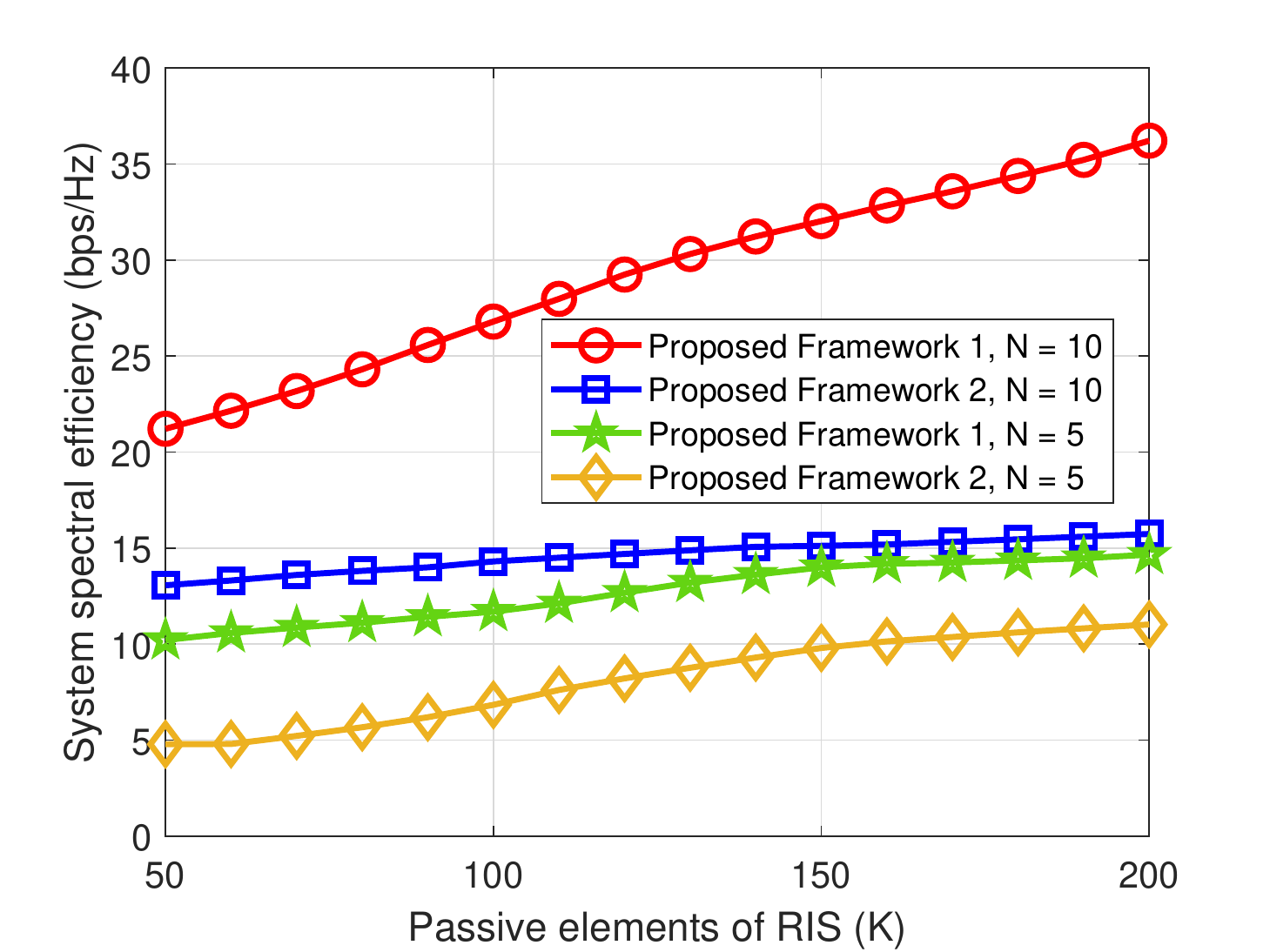}
\caption{The impact of increasing RIS elements on the system spectral efficiency of the proposed framework 1 and proposed framework 2 using different values of $N$.}
\label{Fig2}
\end{figure}

Fig. \ref{Fig2} shows the impact of the increasing number of RIS elements on the system's spectral efficiency. Here, we compare the performance of proposed framework 1 with proposed framework 2. Note that in proposed framework 1, users in each cell receive their signal from BS through both direct and RIS-assisted links. However, in the proposed framework 2, a weak user receives the signal from BS through both direct and RIS-assisted links, while a strong user receives the signal from BS through only a direct link. As expected, increasing RIS elements increases the system's spectral efficiency with different cells for both communication scenarios. It can be seen that we get much better performance when the RIS assists the signal of both users in each cell. All the results in this section showed that the systems with more cells provide very high spectral efficiency. It shows the importance of the proposed scheme in the large-scale multi-cell RIS-NOMA network. 

\begin{figure}[!t]
\centering
\includegraphics[width=0.5\textwidth]{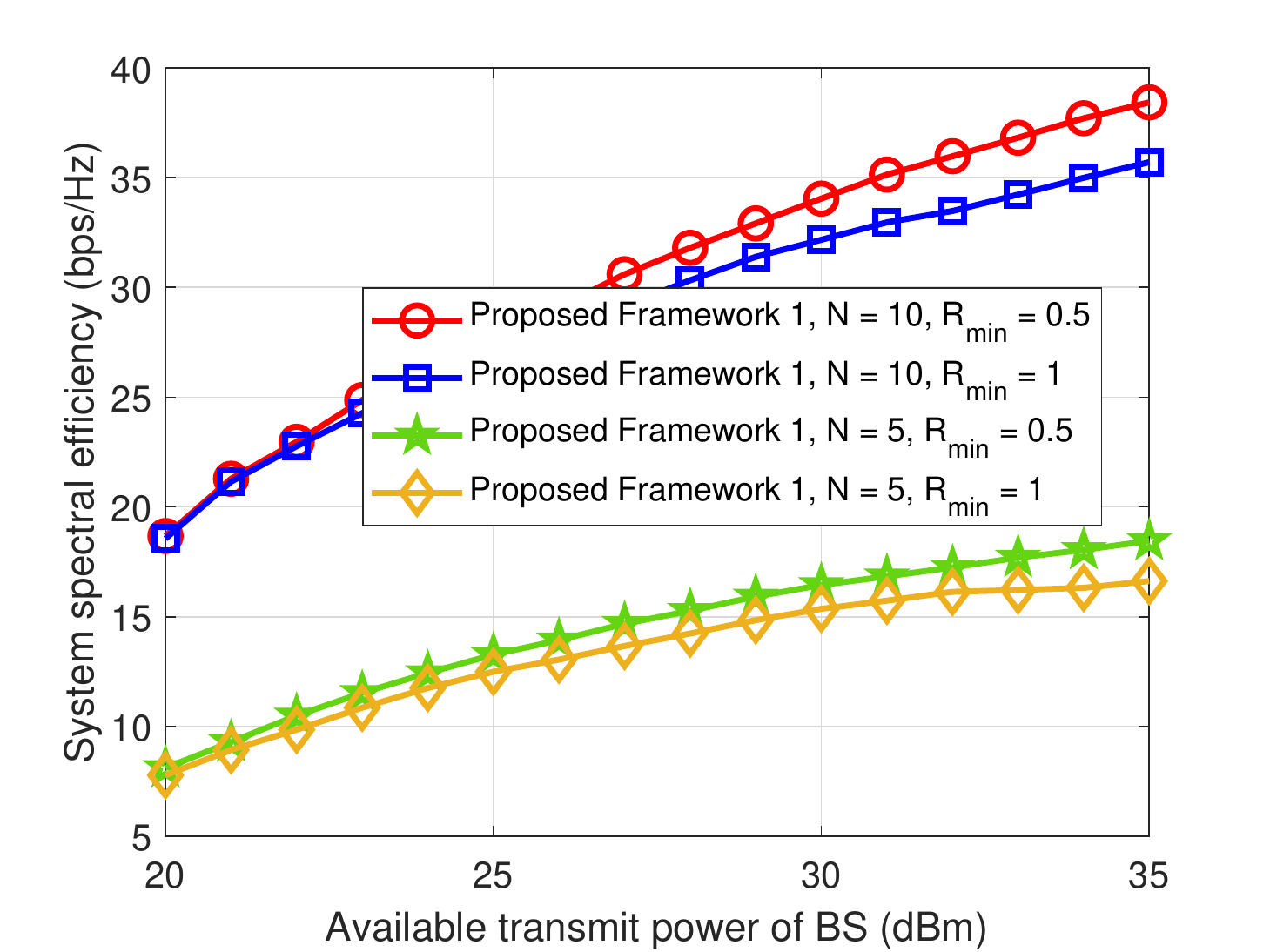}
\caption{The effect of increasing BS power on the system spectral efficiency with different values of $N$ and $R_{min}$.}
\label{Fig3}
\end{figure}

Next, it is important to show the effect of BS allocating power on the system's performance. Fig. \ref{Fig3} shows the increasing values of $P_{tot}$ versus the spectral efficiency of the system. The spectral efficiency increases with increasing the allocated transmit power of BS. Further, the gap in the spectral efficiency offered by all the cases also increases with $P_{tot}$, because more available transmit power at BS becomes available for optimization. Hence, the cases with less restrictions (smaller $R_{min}$ requirements) perform better than other cases with higher rate requirements. Besides that, we can see that the system spectral efficiency similarly increases as the number of cells increases from 5 to 10. It shows the effectiveness of the proposed scheme for a large RIS-NOMA network. 

\begin{figure}[!t]
\centering
\includegraphics[width=0.5\textwidth]{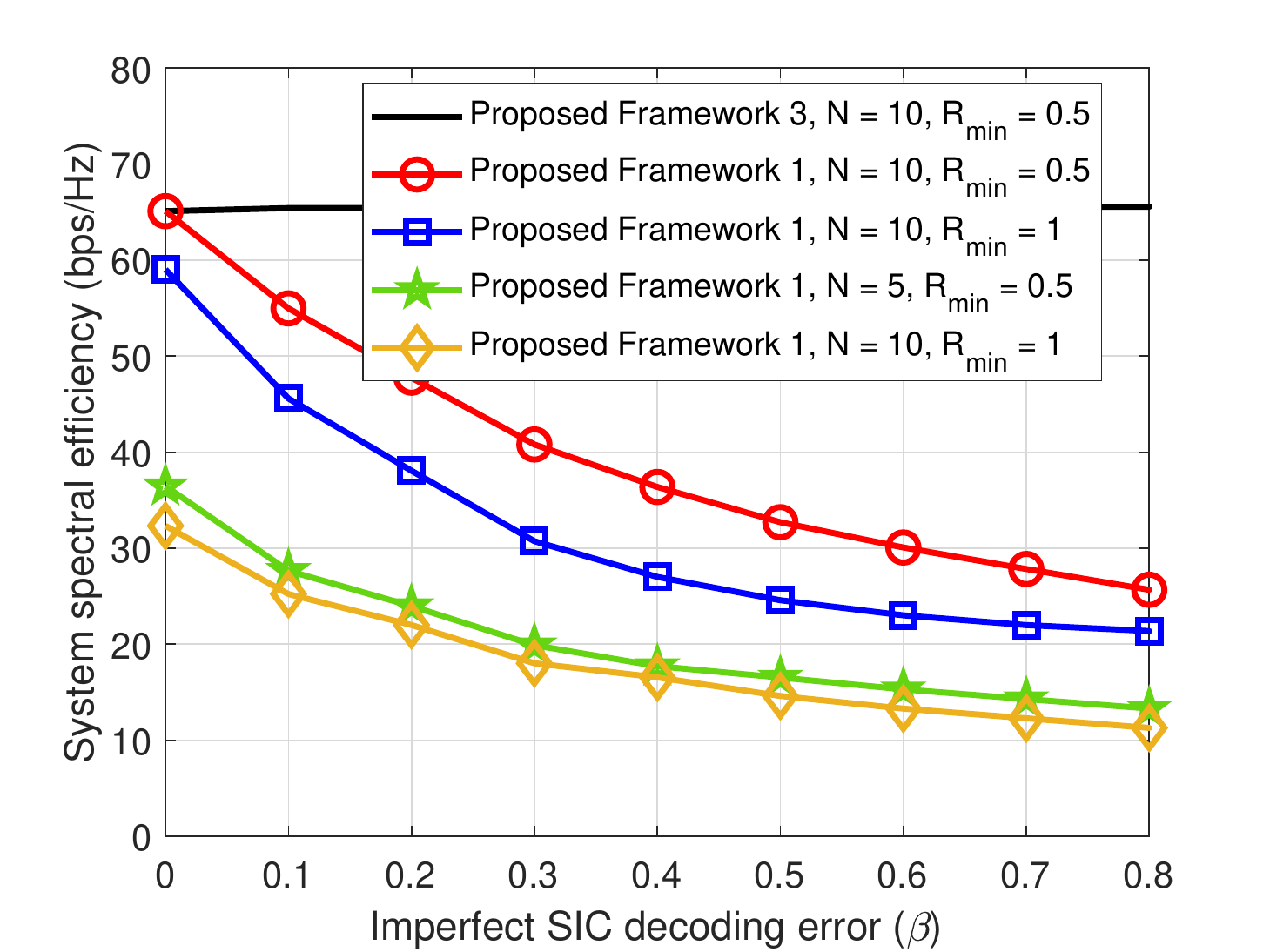}
\caption{The impact of imperfect SIC parameter on the system spectral efficiency of proposed framework 1 and proposed framework 3 with different values of $N$ and $R_{min}$.}
\label{Fig4}
\end{figure}

To check the impact of imperfect SIC decoding on the system performance, Fig. \ref{Fig4} plots the system spectral efficiency against the increasing values of $\beta$ where the number of cells is set as $J=5$ and $J=10$ while the values of $R_{min}$ is 0.5 bps/Hz and 1 bps/Hz, respectively. Here, we compare proposed framework 1 and proposed framework 3. We can see that for all considered cases, the system's spectral efficiency reduces when the value of $\beta$ increases. This is because the high values of $\beta$ increase the interference of NOMA users due to poor signal decoding capability. Another point worth mentioning here is the high decrease in spectral efficiency of those cases with high $R_{min}$. It is because with increasing $R_{min}$, more power is required by users to meet the requirements. Thus, the power allocation becomes less flexible. In practical systems achieving perfect SIC is a difficult task. However, for ease of solution, most of the works in the literature consider perfect SIC. Moreover, as shown in this figure, i.e., proposed framework 3, perfect SIC provides an over-optimistic performance evaluation.

\section{Conclusions}
The integration of NOMA with RIS has the potential to extend wireless communication and connect massive devices in next-generation networks. This letter has proposed a new optimization scheme for a multi-cell RIS-NOMA network to maximize the system's spectral efficiency. Specifically, our framework simultaneously optimizes the transmit power of BS and the reflection matrix of RIS in each cell under SIC decoding errors. The problem of total transmit power of BS and power allocation coefficients of users has been solved using the inner approximation method while the reflection matrix of RIS has been designed by successive convex approximation and DC programming. Simulation results have confirmed the benefits of the proposed optimization scheme. This work can be extended in several ways. For example, the proposed system can be considered using simultaneous transmitting and reflecting RIS, where this work will be treated as a benchmark scheme. This work can also be extended to a multi-antenna scenario, where BS in each cell will be equipped with multiple antennas to communicate with multiple users. In such a system, we can compute active beamforming at BS along with other optimization variables to further enhance the system performance. Last but not least, high-frequency bands such as Terahertz communications can in be achieved for indoor communications in small-cell scenarios. These interesting yet explored communication scenarios and problems will be investigated in our future studies.

\section*{Appendix A: Transformation of $T'_1$ and $T'_2$}
By transforming $T'_1$ of problem \eqref{13}, it can be expressed as:
\begin{align}
R_{i,n}\geq\psi_{i,n},
\end{align}
Writing the value of $R_{i,n}$ as:
\begin{align}
\log_2(1+\gamma_{i,n})\geq\psi_{i,n},
\end{align}
It can Further simplified as:
\begin{align}
\log\left(1+\frac{P_n\alpha_{i,n}\chi_{i,n}}{P_n(1-\alpha_{i,n})\chi_{i,n}\beta+I_{i,n'}+\sigma^2} \right)\geq\psi_{i,n}\log(2),
\end{align}
where $\chi_{i,n}=|g_{i,n}+{\boldsymbol{\delta}_n^H}\text{diag}(\textbf{h}_{i,k,n}) \textbf{f}_{i,k,n}|^2$. Now taking least common multiple (LCM) as:
\begin{align}
&\log\left(\frac{P_n(1-\alpha_{i,n})\chi_{i,n}\beta+I_{i,n'}+\sigma^2+P_n\alpha_{i,n}\chi_{i,n}}{P_n(1-\alpha_{i,n})\chi_{i,n}\beta+I_{i,n'}+\sigma^2} \right)\nonumber\\&\geq\psi_{i,n}\log(2),
\end{align}
Next we apply the property of logarithm, the above equation can be stated as:
\begin{align}
\log\left({P_n(1-\alpha_{i,n})\chi_{i,n}\beta+I_{i,n'}+\sigma^2+P_n\alpha_{i,n}\chi_{i,n}} \right)\nonumber\\-\log(P_n(1-\alpha_{i,n})\chi_{i,n}\beta+I_{i,n'}+\sigma^2)\geq\psi_{i,n}\log(2),
\end{align}
By moving the minus terms to the other side, it can be expressed as:
\begin{align}
\log\left({P_n(1-\alpha_{i,n})\chi_{i,n}\beta+I_{i,n'}+\sigma^2+P_n\alpha_{i,n}\chi_{i,n}} \right)\nonumber\\\geq\psi_{i,n}\log(2)+\log(P_n(1-\alpha_{i,n})\chi_{i,n}\beta+I_{i,n'}+\sigma^2),
\end{align}
Accordingly, we transform $T'_2$ of problem \eqref{13} such as:
\begin{align}
R_{j,n}\geq\psi_{j,n},
\end{align}
Expending the values of $R_{j,n}$, it can be written as:
\begin{align}
\log_2(1+\gamma_{j,n})\geq\psi_{j,n},
\end{align}
It can also be given as:
\begin{align}
\log\left(1+\frac{P_n(1-\alpha_{i,n})\chi_{j,n}}{P_n\alpha_{i,n}\chi_{j,n}+I_{j,n'}+\sigma^2} \right)\geq\psi_{j,n}\log(2),
\end{align}
where $\chi_{j,n}=|g_{j,n}+{\boldsymbol{\delta}_n^H}\text{diag}(\textbf{h}_{j,k,n}) \textbf{f}_{j,k,n}|^2$. Next, by computing LCM, it can be written as:
\begin{align}
&\log\left(\frac{P_n\alpha_{i,n}\chi_{j,n}+I_{j,n'}+\sigma^2+P_n(1-\alpha_{i,n})\chi_{j,n}}{P_n\alpha_{i,n}\chi_{j,n}+I_{j,n'}+\sigma^2} \right)\nonumber\\&\geq\psi_{j,n}\log(2),
\end{align}
Now using the property of Logarithm, we can be expressed as:
\begin{align}
\log\left({P_n\alpha_{i,n}\chi_{j,n}+I_{j,n'}+\sigma^2+P_n(1-\alpha_{i,n})\chi_{j,n}} \right)\nonumber\\-\log(P_n\alpha_{i,n}\chi_{j,n}+I_{j,n'}+\sigma^2)\geq\psi_{j,n}\log(2),
\end{align}
Finally, we move the minus terms to the other side as:
\begin{align}
\log\left({P_n\alpha_{i,n}\chi_{j,n}+I_{j,n'}+\sigma^2+P_n\alpha_{j,n}\chi_{j,n}} \right)\nonumber\\\geq\psi_{j,n}\log(2)+\log(P_n\alpha_{i,n}\chi_{j,n}+I_{j,n'}+\sigma^2),
\end{align}
This end the proof.

\bibliographystyle{IEEEtran}
\bibliography{Wali_EE}

\begin{IEEEbiography}
[{\includegraphics[width=1in,height=1.5in,clip,keepaspectratio]{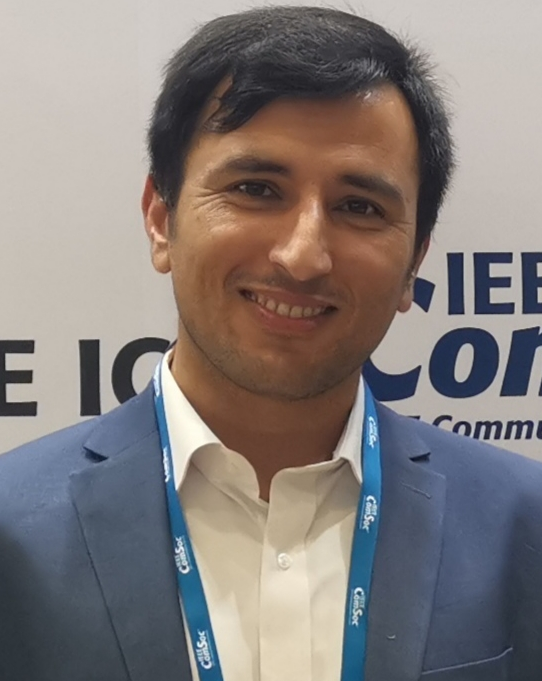}}]{Wali Ullah Khan} (Member, IEEE)
received the Master degree in Electrical Engineering from COMSATS University Islamabad, Pakistan, in 2017, and the Ph.D. degree in Information and Communication Engineering from Shandong University, Qingdao, China, in 2020. He is currently working with the Interdisciplinary Centre for Security, Reliability and Trust (SnT), University of Luxembourg, Luxembourg. He has authored/coauthored more than 100 publications, including international journals, peer-reviewed conferences, and book chapters. His research interests include convex/nonconvex optimizations, non-orthogonal multiple access, reflecting intelligent surfaces, ambient backscatter communications, Internet of things, intelligent transportation systems, satellite communications, unmanned aerial vehicles, physical layer security, and applications of machine learning.
\end{IEEEbiography}

\begin{IEEEbiography}
[{\includegraphics[width=1in,height=1.5in,clip,keepaspectratio]{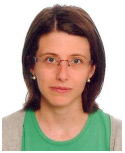}}]{Eva Lagunas} (Senior Member, IEEE)
received the M.Sc. and Ph.D. degrees in telecommunications engineering from the Polytechnic University of Catalonia (UPC), Barcelona, Spain, in 2010 and 2014, respectively. From 2009 to 2013, she was a Research
Assistant with the Department of Signal Theory and
Communications, UPC. In 2009 she was a Guest Research Assistant with the Department of Information
Engineering, University of Pisa, Pisa, Italy. From
November 2011 to May 2012, she held a Visiting
Research appointment with the Center for Advanced
Communications, Villanova University, PA, USA. In 2014, she joined the
Interdisciplinary Centre for Security, Reliability and Trust (SnT), University
of Luxembourg, Luxembourg, where she currently holds a Research Scientist
position. Her research interests include radio resource management and general
wireless networks optimization.
\end{IEEEbiography}

\begin{IEEEbiography}
[{\includegraphics[width=1in,height=1.5in,clip,keepaspectratio]{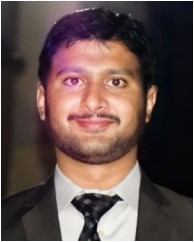}}]{Asad Mahmood}
received the master's degree in electrical engineering from the Department of Electrical \& Computer Engineering, COMSATS University Islamabad, Wah Campus, Pakistan. During his master studies, he has been a Research Assistant with Dr. Yue Hong with the College of Mechatronics and Control Engineering, Shenzhen University, Shenzhen, China. He is currently a PhD student in Interdisciplinary Centre for Security, Reliability and Trust (SnT), University of Luxembourg, Luxembourg. His research interests include resource allocation in wireless communication, mobile edge computing, machine learning, evolutionary algorithm, and classical optimization.
\end{IEEEbiography}

\begin{IEEEbiography}
[{\includegraphics[width=1in,height=1.5in,clip,keepaspectratio]{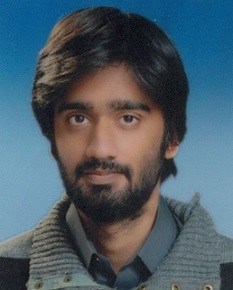}}]{Zain Ali} received the Ph.D. degree in Electrical Engineering from COMSATS University, Islamabad, Pakistan, in 2021. He was awarded HEC’s indigenous scholarship for MS and PhD studies. Currently, he is working as a PostDoc researcher at the Department of Electrical and Computer Engineering, University of California, Santa Cruz, USA. His research interests include cognitive radio networks, energy harvesting, multi–hop relay networks, orthogonal frequency division multiple access (OFDMA), non–orthogonal multiple access (NOMA), Machine Learning, and engineering optimization. 
\end{IEEEbiography}

\begin{IEEEbiography}
[{\includegraphics[width=1in,height=1.5in,clip,keepaspectratio]{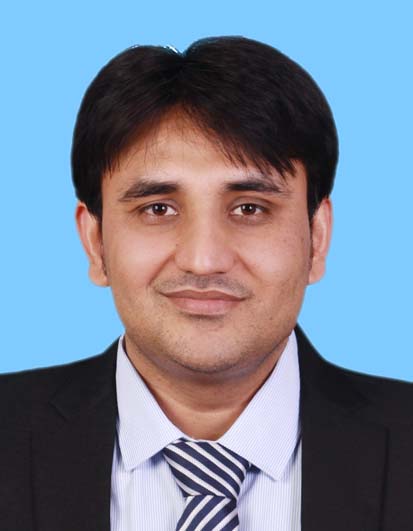}}]{Muhammad Asif} was born in Rahim Yar Khan, Bahawalpur Division, Pakistan, in 1990. He received the Bachelor of Science (B.Sc) degree in Telecommunication Engineering from The Islamia University of Bahawalpur (IUB), Punjab, Pakistan, in 2013, and Master degree in Communication and Information Systems from Northwestern Polytechnical University (NWPU), Xian, Shaanxi, China, in 2015. He also received Ph.D. degree in Information and Communication Engineering from University of Science and Technology of China (USTC), Hefei, Anhui, China in 2019. Currently, Dr. Asif is working as a post-doctoral researcher at the Department of Electronics and Information Engineering in Shenzhen University, Shenzhen, Guangdong, China. He has authored/co-authored more than 25 journal and conference papers. His research interests include Wireless Communication, Channel Coding, Coded-Cooperative Communication, Optimization and Resource Allocation, Backscatter-Enabled Wireless Communication, IRS-Assisted Next-generation IoT Networks.
\end{IEEEbiography}

\begin{IEEEbiography}
[{\includegraphics[width=1in,height=1.5in,clip,keepaspectratio]{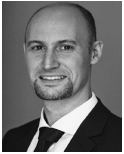}}]{Symeon Chatzinotas} (Fellow, IEEE)
received the M.Eng. degree in telecommunications from the Aristotle University of Thessaloniki, Thessaloniki, Greece, in 2003, and the M.Sc. and Ph.D. degrees in electronic engineering from the University of Surrey, Guildford, U.K., in 2006 and 2009,
respectively. He is currently a Full Professor or Chief
Scientist I and the Co-Head of the SIGCOM Research Group, Interdisciplinary Centre for Security, Reliability and Trust, University of Luxembourg. In the past, he was a Visiting Professor with the University of
Parma, Parma, Italy, and he was involved in numerous Research and Development projects for the National Center for Scientific Research Demokritos, the Center of Research and Technology Hellas and the Center of Communication
Systems Research, University of Surrey. He has coauthored more than 400
technical papers in refereed international journals, conferences and scientific
books. He was the co-recipient of the 2014 IEEE Distinguished Contributions
to Satellite Communications Award, the CROWNCOM 2015 Best Paper Award,
and the 2018 EURASIP JWCN Best Paper Award. He is currently in the
Editorial Board of the IEEE OPEN JOURNAL OF VEHICULAR TECHNOLOGY and
the International Journal of Satellite Communications and Networking.
\end{IEEEbiography}

\begin{IEEEbiography}
[{\includegraphics[width=1in,height=1.5in,clip,keepaspectratio]{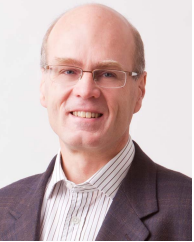}}]{Bj\"orn Ottersten} (Fellow, IEEE)
was born in Stockholm, Sweden, in 1961. He received the M.S. degree in electrical engineering and applied physics from Link\"oping University, Link\"oping, Sweden, in 1986, and the Ph.D. degree in electrical
engineering from Stanford University, Stanford, CA, USA, in 1990. He has held Research positions with
the Department of Electrical Engineering, Linkoping University, Linkoping, Sweden, the Information Systems Laboratory, Stanford University, the Katholieke Universiteit Leuven, Leuven, Belgium, and the University of Luxembourg, Esch-sur-Alzette, Luxembourg. From 1996 to 1997,
he was the Director of Research with ArrayComm, Inc., a Start-Up in San
Jose, CA, USA, based on his patented technology. In 1991, he was appointed
Professor of signal processing with the Royal Institute of Technology (KTH),
Stockholm, Sweden. He has been the Head of the Department for Signals,
Sensors, and Systems, KTH, and the Dean of the School of Electrical
Engineering, KTH. He is currently the Director for the Interdisciplinary Centre
for Security, Reliability and Trust, the University of Luxembourg. He was the
recipient of the IEEE Signal Processing Society Technical Achievement Award
and been twice awarded the European Research Council Advanced Research
Grant. He has coauthored journal papers which was the recipient of the IEEE
Signal Processing Society Best Paper Award in 1993, 2001, 2006, 2013, and
2019, and eight IEEE conference papers best paper awards. He has been a
Board Member of IEEE Signal Processing Society and the Swedish Research
Council and currently serves on the Boards of EURASIP and the Swedish
Foundation for Strategic Research. He was an Associate Editor for the IEEE
TRANSACTIONS ON SIGNAL PROCESSING and the Editorial Board of
the IEEE Signal Processing Magazine. He is currently a Member of the
Editorial Boards of the IEEE OPEN JOURNAL OF SIGNAL PROCESSING,
EURASIP Signal Processing Journal, EURASIP Journal of Advanced Signal
Processing and Foundations and Trends of Signal Processing. He is a Fellow
of EURASIP.
\end{IEEEbiography}

\end{document}